\documentclass[12pt]{article}

\usepackage[font=footnotesize]{caption}
\usepackage{afterpage}

\usepackage{mathptmx}
\usepackage{multirow}
\usepackage{graphicx} 
\usepackage{pgfornament}
\usepackage{psvectorian}
\usepackage{framed}

\usepackage[letterpaper,margin=1in]{geometry}

\usepackage{setspace}
\usepackage[tiny,compact]{titlesec}
\titleformat*{\section}{\large\bfseries}

\usepackage{url}
\usepackage{amsmath,amsthm,amssymb,textcomp,mathrsfs,cancel,bm}
\usepackage{graphicx}
\newtheorem{thm}{Theorem}[section]
\newtheorem{cor}[thm]{Corollary}
\newtheorem{postulate}[thm]{Postulate}

\usepackage{xhfill}

\usepackage{verse}

\usepackage{etoolbox,bigints}
\usepackage{lipsum}

\renewcommand\qedsymbol{$\blacksquare$}

\newcommand\blankpage{%
    \null
    \thispagestyle{empty}%
    \addtocounter{page}{-1}%
    \newpage}

\theoremstyle{definition}
\newtheorem{innerexample}{Theorem}

\makeatletter
\patchcmd{\endinnerexample}{\endtrivlist}{\endlist}{}{}
\newenvironment{example}
 {\patchcmd{\@thm}{\trivlist}{\list{}{\leftmargin=3em \rightmargin=3em}}{}{}%
 \innerexample\pushQED{\qed}}
 {\popQED\endinnerexample}
\makeatother

\newcommand{\centereqn}[3]
{
\fbox{
	\begin{minipage}[#1]{#2}
        #3
	\end{minipage}
	}
}

\usepackage{pdfpages}

\begin{document}



# General Solution For Generalised Newell–Whitehead–Segel Equations

author

**Luisiana X. Cundin**

**August 05, 2020**

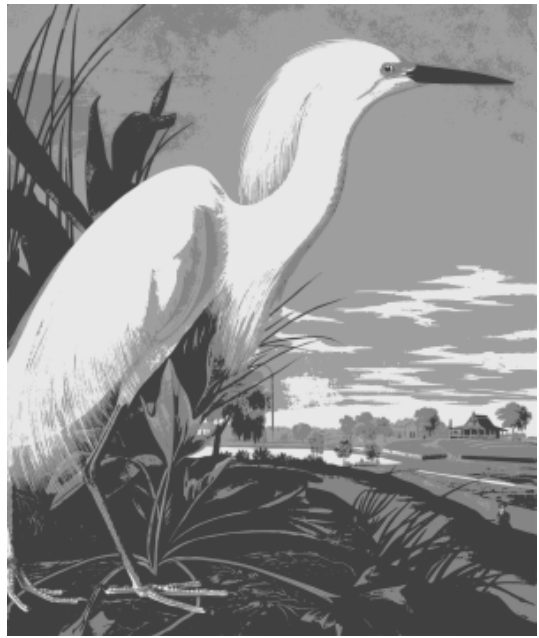

Plate 242, "Snowy Heron", or "White Egret," Audubon Society,
https://www.audubon.org/birds-of-america/snowy-heron-or-white-egret

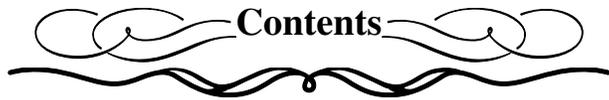
# Contents

# Abstract

In this monograph, two sets of parabolic differential equations are studied, each with nonlinear medium response. The equations are generally referred to as "Newell–Whitehead–Segel equation," which model a wide variety of nonlinear physical, mechanical and biological systems. Nonlinear medium response can be viewed in many perspectives, such as, memory response from the medium, whereby, the medium "remembers" earlier influences; reactive responses, whereby, the medium is actively responsive to input, such as, chemical reactivity, turbulence and many other circumstances; these equations arise often in the biological sciences when modeling population dynamics, whether the population be genomic, such as, alleles, or animal species in the environment, finally, these sets of equations are often employed to model neurological responses from excitable cellular media.

The solutions provided are of a very general nature, indeed, whereby, a canonical set of solutions are given for a class of nonlinear parabolic partial differential equations with nonlinear medium response expressed as either a $p$–times iterative convolution or $p$–times multiplicative response. The advantage of canonical solution sets are these solutions involve classic representative forms, such as, Green's function or Green's heat kernel and aid researchers in further complication, analysis and understanding of the systemic behavior of these types of nonlinear systems.

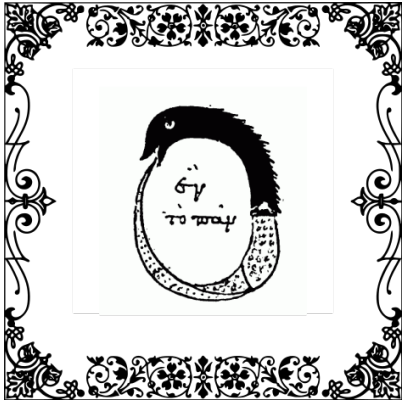

Ouroborus

---

**keywords:** nonlinear parabolic partial differential equations, Fisher's equation, Nagumo equation



# 1 Preface

The topic of nonlinear partial differential systems is voluminous...so are the methods employed, both analytic and numerical techniques alike. Attempting to solve these rather intractable problems, most approaches fall to numerical solutions and approaches in nature, because of the complex nature of these equations and systems. Any approach to nonlinear systems requires a focused perspective on a small class of functions, usually, with fixed order of nonlinearity; otherwise, the analyst is swiftly overwhelmed by all possible combinations. Fortune, on occasion, can lend the rarest of opportunities where a set of generalized nonlinear partial differential equations admit solution; such is the case in this monograph.

Linear systems, more or less, enable analytic solutions to be obtained, and, the benefit of analytic solutions cannot be understated. The crown–jewel of any analytic solution is what is referred to as a Green's function, which represents the most fundamental impulse response of the system in question. If such an impulse solution were possible for a given nonlinear system, then enormousness advantages would be reaped from possessing such irreducible solutions in the form of a Green's kernel solution.

Generally speaking, the Newell–Whitehead–Segel equation is used to model reaction–diffusion systems. Nonlinear systems span the spectrum of hard Sciences: from physics, chemistry to biology. Notwithstanding the immense intricacies and delight for any given application, in general, peculiarities for any one application would inevitably involve numerous parts, circumstances and factors, which make analysis for any given specific problem, say, in tracing the prominence of a genetic allele propagating through a population of animals, quite a story in–and–itself; therefore, the purpose of this monograph is by far not to address all such specific applications, rather, the purpose of this monograph is to show and demonstrate a rather general methodology for finding solutions to a specific class of nonlinear parabolic differential equations.

My graduate advisor, Dr. Ashok Puri, University of New Orleans, Physics Depart, asked me to solve a specific nonlinear parabolic differential equation for my dissertation. Since the advent of the coronavirus pandemic, I have been forced to walk away from university; yet, the work continued and matured. It is my expressed ambition and desire to aid further researchers, especially, young researchers in the field of mathematical physics by detailing said techniques as outlined in this monograph; moreover, it is my strong conviction these techniques hold deeper significance and possess a wider application, which could greatly aid young researchers seeking to solve yet further obstreperous and refractory mathematical equations.

Kind Regards,

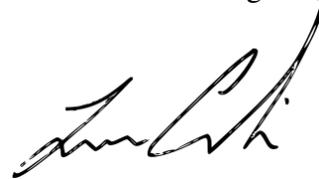

Luisiana Cundin
July 19, 2020



    \setcounter{page}{4}

\section{Prop\ae deutics}

Nonlinear dynamic systems both fold and stretch the experimental space within which they reside; repeated folding and stretching ultimately leads to chaotic behavior. Chaos is of particular interest to many modern researchers in their study of complex dynamic systems. Chaotic behavior is a deterministic system at heart, yet behaves seemingly random over time; thus, revealing a manifold world  that enjoys an array of very rich and convoluted behavior. Even though the ''heat equation'' is widely perceived a dynamic process not exhibiting chaotic behavior, this most fundamental process describes some of the most elementary physical processes in the universe, namely, information content of a system, which physicist also deem as entropy. Chaos becomes evident within heat evolution if the the heat equation is run in reverse, the so\textendash called ''Reversed Heat Equation,'' whereby, time is reversed and cooler regions pool and pile heat upon any hotter region in the vicinity. The implosion of fissile material exhibits this behavior, where unintended hot spots form within the kernel of the plum, causing unstable, inconsistent chain reactions. Ultimately, in reverse, the heat equation does exhibit chaotic behavior, thus, despite being unaware, this most fundamental physical process is chaotic in both directions.

Of course, focusing on the linear heat equation, the equation simply describes the flow of heat from hot regions to colder regions in some domain $\mathcal{D}$. This natural phenomenon may appear, at first, quite unexciting, nevertheless, heat is ultimately subject to the rules of quantum mechanics, therefore, the entire process is probabilistic in nature. In fact, the heat equation, proper, essentially describes the statistical average or aggregate behavior of a system seeking equilibrium, for example, Einstein's description of Brownian motion is the heat equation, basically describing the aggregate, random motion of Brownian particles \cite{Einstein}. The classic example of chaos is the Lorenz systems, which describe atmospheric convection, but all air currents in the atmosphere are governed by temperature; therefore, even this process is ultimately ruled by diffusion\textendash convection processes. Another example, in solid state physics, heat is treated as a quantum particle, referred to as a phonon, and is described by Schr\"{o}dinger's equation, which is the heat equation rotated in the complex plane to reflect the conjugate solution on Riemann's sphere. In quantum mechanics, the basic solutions, so\textendash called wave functions, are always time averaged and spatially averaged, therefore, once again, the aggregate behavior constitutes what is observable. Even in pure mathematics examples exist for the utility of the heat equation, the so\textendash called Ricci flow is a mathematical method to explore topological surfaces \cite{Ricci}. Examples abound, but for one additional example, diffusion processes are widely used in the biological sciences to model everything from cellular migration, biological cellular growth, genomics, mutagenesis, population dynamics, and the list goes on\ldots

There are few analytic solutions for nonlinear equations, thus, analyst resort to numerical analysis to study and generate solutions for such models. There exists many prohibitive numerical reasons in attempting a study of long\textendash time chaotic behavior for a nonlinear system, mainly, the nature of computers and how they represent numbers in binary form disallow continuous number representation \cite{Computer}. An algorithm run on a computer can become ''stuck in a rut,'' so to speak, producing false sets of solutions. The number line is not equally spaced along the binary representation in the IEEE numbering system, hence, round\textendash off error causes interlaced sampling problems (\emph{see Bracewell for details and implications}), plus, error accumulates over the run of the algorithm, leading to biased results \cite{Bracewell}. This fact is unfortunately true for any numerical method entertained: finite difference, finite element, finite volume, interpolation, \&c; as a consequence, there is no substitute for analytic solutions for differential systems, in general. Closed\textendash form, analytic results provide a global solution for analysts, whereby, in\textendash depth analysis may be performed for a system under study; in addition, analytic results provide a kernel for further complications, which can either yield additional closed\textendash form results, but, if not, then enabling and aiding analysts in minimizing numerical error under further integration or under some other mathematical operation.

The differential systems studied in this monograph are quite general in nature and provide closed\textendash form, analytic solutions for a class of nonlinear parabolic differential equations. These equations are called by many names, but are known as the ''Newell\textendash Whitehead\textendash Segel equation,'' which has the following general form:
\begin{equation*}
 \frac{\partial}{\partial t}u(x,t)- D\frac{\partial^2}{\partial x^2}u(x,t)+b\, u(x,t)-\epsilon\, f(u)=0,
\end{equation*}
\noindent where $f(u)$ is functional of the unknown function $u(x,t)$.

The differential system is governed by a set of constants $\{D,b,\epsilon\}$, where $D$ is the diffusion rate or constant, $b$ is the convection force and $\epsilon$ is the magnitude of the nonlinear medium response, which is represented by the functional $f(u)$. For the linear case, \emph{inept}, for $\epsilon$ and $b$ equal to zero, the equation is referred to as the heat equation. Solutions abound for the linear case, but the most fundamental solution is that referred to as Green's solution or Green's heat kernel, \emph{viz.}:
\begin{equation*}
 G(x,t)=\frac{e^{-\frac{x^2}{4Dt}}}{\sqrt{4\pi Dt}}
\end{equation*}

Green's solution represents the most fundamental solution because it is irreducible. All other solutions, in any other form or representation, is either Green's heat kernel or an amalgamation of Green's heat kernel with other functions. It is most advisable and satisfying to achieve or find a Green's solution for a given differential system, because it provides the most basic of solutions. If further complications are entertained, such as, a multi\textendash regional domain with differing diffusion constants, then a global solution can be constructed by suitable choice of boundary conditions for each region, thereby, forming a quotient space to describe the entire domain $\mathcal{D}$. Green's heat kernel reduces to a Dirac delta function of space in the limit of zero time, thus, describing one point in the entire domain as the input of heat for all later time, hence, Green's heat kernel simply represents one point in space initially heated and then describes how this energy diffuses or spreads out over time. Having Green's function reduce to a Dirac delta function is crucial, because, if further complication be desired, it is a simple matter of convolving whatever spatial distribution under consideration with Green's heat kernel to arrive at that solution. 

This property is generally referred to as a transfer function, because the kernel is an impulse function. In general, further complications can be achieved by convolving a driving function $f(x,t)$ with Green's heat kernel, thus:
\begin{equation*}
 H(x,t)=\iint{f(x-x',t-t')G(x',t')dx'dt'}
\end{equation*}

Numerically speaking, this property carries great weight, because the kernel carries the core, key and the bulk of the information for the system, therefore, Pareto is achieved by simply studying Green's solution; but, in addition, since the majority of the information is held in hand as an exact solution, then all further complications can mitigate numerical error, aid in forming complex quotient spaces and a host of other mathematical operations. 

If convection be considered, \emph{inept}, $b$ is not equal to zero, then the solution requires an exponential mapping from Green's solution to a plane tangent to Riemann's sphere, namely, Green's function would be multiplied by the exponential of the convection constant multiplied by time, i.e. $G(x,t)e^{-bt}$. The exponential mapping is the result of integrating the convection constant over time and can be considered a winding number, thus, the larger the convection, the faster decay of the system; in other words, think of heat in a convective system diffusing and being further carried by fluid flow. Of course, if the convection constant is negative, then the convection process is reversed, heat is being driven into the system, therefore, the solution would grow exponentially in time. 

Finally, considering the nonlinear medium response, which is governed by the constant $\epsilon$, there is an array of possible responses one can consider; ultimately, the system is now a function of a set of constants and the function itself, i.e. $\{D,b,\epsilon,f(u)\}$. A more formal, general definition of the system's dependence would be a functional with a set of variables that may or may not be time or spatially dependent, in addition, the system itself plays a role in deciding the behavior at any given moment in the experimental space, \emph{viz.}:
\begin{equation*}
 F(x_i,D,b,\epsilon,u;x_i,t)=0
\end{equation*}

The functional $F$ is a most general functional, whereby, all the independent variables could be a function of the spatial variables, which could be themselves a set of independent variables indexed by $i$. Of course, all variables could be indexed to reflect a domain comprised of multiple regions ($R_i$), i.e. $\sum R_i\subset\mathcal{D}$. Finally, all independent variables could be time dependent. The functional considered is homogeneous and this will be the only circumstance considered in this monograph; although, due diligence is given for further complications in the case of inhomogeneous problems. 

There are two fundamental medium responses to be considered in this monograph, namely, a nonlinear medium response where the functional $f(u)$ is either an iterated convolution or multiplicative response. Now, non\textendash linearity is a process of folding and stretching space and this leads to chaotic behavior; in both functionals considered, the basic process of folding and stretching of space is involved. In the case of convolution, generally, two functions are multiplied together and integrated over a shifted variable; this process tends to smooth or stretch the space. In like manner, multiplication of two functions is equivalent to the convolution of the Fourier transform of each function, hence, the same process is occurring, only in the frequency domain. 

The first case to be considered will be the iterated convolution of the medium and may be thought of as the medium possessing ''memory,'' whereby, information is folded from the past onto the present. In probability theory, the distribution of a set of random, independent variables are added, the normalized sum tends to a normal distribution, even if the original distribution was not normally distributed, i.e. the \emph{Central Limit Theorem}. This is evidenced by Green's heat kernel, which is certainly Gaussian in spatial terms, thus, the spatial average of the heat particles, Brownian particles, \&c, tend to a smooth, Gaussian distribution. Of course, the Central Limit Theorem seeks tendencies to a Gaussian distribution, \emph{inept}, a normal distribution $\mathcal{N}(\mu,\sigma)$; nonetheless, seeking beyond this limit would find the ultimate conclusion of an infinite convolution to tend to unity, in other words, a normal distribution, whose variance is ever widening. 

Contrariwise, another approach would be to consider a multiplicative medium response, which is predominately a folding of the space, for example, consider the following:
\begin{equation*}
 x^2-1=0
\end{equation*}

The order of multiplication governs the number of solutions or roots of the system and represents the experimental space, essentially, being folded upon itself, where the solutions reside on two separate Riemann sheets. In the extreme, the logarithmic function with complex variable yields an infinite set of solutions, $\log(z)+i(\theta+2k\pi)$, where each solution resides on separate Riemann sheets, indexed by integer $k$. Consequently, this is the reason for seeing analyst seeking various roots to nonlinear equations, usually, some derived algebraic eigenvalue, and, once solved, enables the analysts to parse out specific solution sets on the complex domain $\mathbb{C}^n$. Brute force methods of identifying various algebraic roots comes with frequent frustrations and failures, worse, the solution centered on may only be valid within a confined range of the complex phase, hence, leading to many unruly, interval dependent solutions that involve such functions as inverse trigonometric functions. 

A multiplicative functional could represent a joint set of independent probabilities, whereby, probabilities are multiplied together; this is often the perception in the biological sciences. In the physical sciences, multiplicative medium responses represent a reactive media, such as, in chemistry, where an aliquot of chemical species react over time; either endo/exothermic in nature. Other examples abound; nonetheless, primarily, the process under study is multiplicative in nature, that is to say, the probabilities are independent. 

On the contrary, as will be shown, the general solutions for either the multiplicative case or the iterated convolution problem are quite general in nature, but the solution can involve some rather ornery special functions, such as,  the incomplete gamma functions or the exponential integral. Both representations indicate the nature of the global solution involved, being nonlinear in nature, time is not completely separable from the system's solution, thus a pole exists somewhere in the complex plane, often residing on the origin. 

The very nature of nonlinear differential systems are self dependent or self\textendash feeding; therefore, like an ouroboros eating its own tail, the system is constantly subject to a continuous feedback loop. Solutions evade capture by analysts because no beginning or end presents itself, thus, it is necessary to catch the dragon by its tail and feed its tail to its mouth to initiate the nonlinear process.

The solutions are all solved and presented in the Fourier codomain, and, for two important reasons: firstly, transformation to the codomain enables finding the exponential time mapping, which would be generally impossible if the spatial variable were explicitly involved; secondly, given the nature of the poles, their movements in time, \&c, inversion from the codomain is generally prevented for either system considered in this monograph.

All differential equations can be perceived as intimately associated with the theory of distributions and  from this point of view, the solutions outlined in this monograph have far reaching implications. The range of possible functions to entertain are far too large for one individual, therefore, it is my hope to foster further curiosity and future research into this area, moreover, it is my expressed hope that the methodology outlined should greatly aid young researchers in his/her quest. 

The methodology employed in this monograph takes advantage of several properties convolutions possess, which, judiciously applied, enable separating the exponential mapping to the Bernoulli differential equation. Once the equation is reduced to the Bernoulli differential equation, an exact solution is admitted for these nonlinear ordinary differential equations. Ultimately, my method is based upon the belief: 
\vspace{7pt}
\noindent\begin{center}\begin{minipage}{0.85\textwidth}
\begin{postulate}{Green's function represents an irreducible solution to the linear differential system; therefore, it is postulated: any further complication for the nonlinear differential system is best constructed from a set of Green's functions, thus, generating a solution comprised of canonical sets of functions.}
\end{postulate}
\end{minipage}\end{center}

\vspace{7pt}

\noindent\begin{center}\begin{minipage}{0.85\textwidth}
\begin{postulate}{The exponential mapping is primarily independent of the base functions involved and dependent upon the nonlinear functional alone.}
\end{postulate}
\end{minipage}\end{center}

\vspace{7pt}

Before closing, I feel it remiss if I did not mention that many would find the inability to transform the solution back into the original domain, that is to say, unsuccessfully finding the inverse Fourier transforms for the solutions as deficient. In my opinion and experience, the spectral domain offers far greater freedom; moreover, after careful consideration, the codomain should be considered superior to the original domain. I would venture to say that most readers of this monograph would like to see inversion, if for no other reason, than to enable ease in applying boundary conditions; but, it is all together possible to apply boundary conditions in the spectral domain, therefore, the pressure or need for inverting the solutions simply does not exist, at least, not in my mind. 

With the advent of computers, it is a simple matter to generate the requisite lists for a discrete Fourier inversion and, thereby, obtain the same. A generalized inversion of the solutions presented in this monograph are, I believe, without any analytic representations, especially, when considering how the complex plane would be broken into multiple branches by the index of non\textendash linearity \textit{p}. This last point is quite prohibitive in attacking a generalized Fourier inversion of the solutions provided. It is altogether possible specific inversions be found, but it would require defining the constant of integration, the power of non\textendash linearity, $p$, and extraordinary work in complex integration. 

Additional benefits exist working solely in the frequency domain, for example, if the coefficients are spatially dependent, then multiplication in the spectral domain can realize this fact; similarly, as seen in electromagnetic applications involving wave propagation and complex index of refraction, whose dependence is both, possibly, spatially and temporally dependent. For stochastic systems, the variables would more easily be treated in the spectral domain. I could cite additional benefits for adhering to the spectral domain, nevertheless, I'll leave the discussion here.

\section{Nonlinear Convolutional Dissipative Systems}

Consider the following homogeneous parabolic partial differential equation, where medium response is nonlinear, expressed as \emph{p}-times, self\textendash repeated convolution, \emph{viz.}:
\begin{equation}
\begin{split}
 \frac{\partial}{\partial t}u(x,t)-& D\frac{\partial^2}{\partial x^2}u(x,t)+b\, u(x,t)-\epsilon\, \Big(u(x,t)\ast u(x,t)\ast\cdots\ast u(x,t)\Big)=0\\
  \{ x | x \in \mathbb{R} \},\, & \{ t | t \in [0,\infty) \},\, \{ \{D,b,\epsilon\} | \{D,b,\epsilon\} \in \mathbb{R}  \}   , \{ p | p \in \mathbb{Z}^{\neq} \text{ and } \mathbb{Z} >1  \} 
 \end{split}\label{diffeqconv}
\end{equation}

Assuming the general solution $u(x,t)$ to the partial differential system is comprised of a convolution of two functions, namely, $u(x,t)=G(x,t)\ast f(x,t)$, where $G(x,t)$ is Green's function\textemdash a known solution for the linear partial differential equation\textemdash a series of particular advantages can be capitalised regarding the properties of convolutions. The first convolution property enables distributing a derivative not explicitly involved in the convolution, resulting in a standard operation for a derivative on a set of functions, see Theorem (\ref{TH1}) in Appendix (\ref{Appendix}). The second property of convolutions to be capitalised is the arbitrary choice to wit a derivative is applied to in a convolution, where the derivative is over the variable of the convolution, see Theorem (\ref{TH2}) in Appendix (\ref{Appendix}).

Employing the above two Theorems enables reducing the original differential equation to the Bernoulli differential equation, \emph{viz.}:
\begin{equation}
\begin{split}
  gF' +b\, Fg- & \epsilon\, F^p g^p=0\\
  \{ s | \{x\mapsto s\} \in \mathbb{C} \},\,  \{ t | t \in [0,\infty) \},\, & \{ \{D,b,\epsilon\} | \{D,b,\epsilon\} \in \mathbb{R}  \} , \{ p | p \in \mathbb{Z}>1  \} \\
 u(x,t)=  \Big(f(x,t)\ast G(x,t)\Big) & \subset F(s,t)g(s,t),
 \end{split}\label{diffeqconvf} 
\end{equation}
\noindent where the convolution is in the spatial domain, employing Bracewell's symbol for Fourier transformations ($\subset$), and, prime ($'$) indicates the derivative with respect to time \cite{Bracewell}.

\begin{center}
 \begin{minipage}{.75\textwidth}
\begin{proof}

 Since Green's function is a known solution for the linear differential equation, its time derivative cancels the second order spatial derivative; thus, after substituting the convolution for the unknown function, $u(x,t)$, the original equation, eq. (\ref{diffeqconv}), can be simplified with the aid of theorem (\ref{TH1}) and theorem (\ref{TH2}), \emph{see Appendix (\ref{Appendix}}), \emph{viz.}:
 \begin{equation*}
  f'\ast G+ \underbrace{f\ast G'-D\Big(f\ast G_{xx}\Big)}_{\mathrm{cancels}}+b\, \Big(f\ast G\Big)-\epsilon\, \Big(f\ast G\ast f\ast G\ast\cdots\ast f\ast G \Big) = 0
 \end{equation*}
 
 Finally, using the property that the Fourier transform of a convolution equals the multiplication of each transformed function, theorem (\ref{TH3}): Appendix (\ref{Appendix}), the original equation is reduced to the Bernoulli differential equation:
 \begin{equation*}
  F'g+bFg=\epsilon\, F^pg^p,
 \end{equation*}
 \noindent which is equivalent to equation (\ref{diffeqconvf}).
 
 \vspace{7pt}
 
 Done.
\end{proof}
\end{minipage}

\end{center}

The Bernoulli differential equation is a nonlinear ordinary differential equation, and, admits exact solutions, typically solved by employing the integrating factor method; consequently, a solution is possible for the reduced differential equation, eq. (\ref{diffeqconvf}), therefore, a solution exists for the original differential equation, eq. (\ref{diffeqconv}), \emph{viz.}:
 \begin{equation}\label{convansw}
\begin{split}
 u(s,t)=& g(s,t)F(s,t)= gh^m= g\, h^{-\left(\frac{1}{p-1}\right)}, \\
 h(s,t)= & e^{b(p-1)t}\Big(C(s)-(p-1)\epsilon\, \int{g^{(p-1)}e^{-b(p-1)t}dt}\Big),
 \end{split}
 \end{equation}
 \noindent where $C(s)$ is the constant of integration resulting from integration over time.
 
%\begin{equation}\label{eq4}
%\begin{split}
% u(s,t)=& g(s,t)F(s,t)= gh^m= g\, h^{-\left(\frac{1}{p-1}\right)}, \\
% h(s,t)= e^{b(p-1)t}\Big(1+(p-1)\epsilon\, & \int{g^{(p-1)}e^{-b(p-1)t}dt}\Big\vert_{t=0}-(p-1)\epsilon\, \int{g^{(p-1)}e^{-b(p-1)t}dt}\Big)
% \end{split}
% \end{equation}
 
The coefficient of integration $C(s)$ is represented in the codomain and completely arbitrary. The second term in function $h(s,t)$ is the general nonlinear kernel solution, which represents the $(p-1)$ multiplication of Green's function in Fourier space, integrated with the integrating factor over time. The time integral is indefinite, but could be made definite for specificity, since, the time domain is defined to be $\{t|t\in\mathbb{R}\geq 0\}$.

Let's show the general solution to the Bernoulli differential equation satisfies the general requirements, \emph{viz.}:

\begin{center}
 \begin{minipage}{.75\textwidth}
\begin{proof} If equation (\ref{convansw}) is a solution for equation (\ref{diffeqconvf}), then taking the derivative with respect to time yields:
 \begin{equation*}
  \frac{\partial}{\partial t}u(s,t)=g'\, F+g\, F'=g'\, F-\frac{g}{p-1}\, h^{-\left(\frac{p}{p-1}\right)}h',
 \end{equation*}
 \noindent where the first term cancels the second order spatial derivative of Green's function, (\emph{as already discussed}). It is only a matter to prove the second term satisfies equation (\ref{diffeqconvf}), therefore, consider the following definitions:
 \begin{align*}
  h'=& b(p-1)e^{b(p-1)t}\, h-(p-1)\epsilon\, g^{p-1}\\
  F'g=-\frac{g}{p-1}\, & h^{-\left(\frac{1}{p-1}\right)}\, h^{-1}\, h'\equiv -bgF+\epsilon\, g^pF^p
 \end{align*}

 Recognize $F(s,t)=h^m$, and, parameter $m=-1/(p-1)$, hence, raising the function $F(s,t)$ to the $p^{\text{th}}$ power yields the identity:
 \begin{equation*}
   F^p=(h^m)^{p}=h^{mp}=h^{-\left(\frac{1}{p-1}\right)-1}=h^{-\left(\frac{p}{p-1}\right)}
  \end{equation*} 

Equivalence has been shown.

 \vspace{7pt}
 
 Done.
\end{proof}
\end{minipage}

\end{center}

\noindent\rule{\textwidth}{3pt}

\begin{example}[\textbf{General solution for the convolutional Newell\textendash Whitehead\textendash Segel equation}]\label{theoremconvsoln}

A most general solution has thus been obtained, representing a closed\textendash form, analytic solution for a homogeneous nonlinear parabolic differential equation, whose nonlinear medium response is expressed as a \textit{p}-iterated convolution, viz.:
\begin{equation*}
 \frac{\partial}{\partial t}u- D\frac{\partial^2}{\partial x^2}u+b\, u-\epsilon\, \Big(u\ast_p\Big)=0 
\end{equation*}

The solution is comprised of Green's function $G(x,t)$ and a nonlinear kernel function $f(x,t)$, whose inverse Fourier transform is defined to be the inverse of the function $h(s,t)$ raised to the power of parameter $m=-1/(p-1)$, \emph{viz.}:
\begin{equation}\label{gensol1}
 u(x,t)= G(x,t)\ast f(x,t)= G\ast\, \mathscr{F}^{-1}\Bigg\{h^{-\left(\frac{1}{p-1}\right)}\Bigg\},
\end{equation}
\noindent where the inverse Fourier transform is signified by symbol $\mathscr{F}^{-1}$, lastly, function $h(s,t)$ is defined as such:
\begin{equation*}
 h(s,t)= e^{(p-1)bt}\Big(C(s)-\epsilon\, (p-1)\int{g^{(p-1)}e^{-b(p-1)t}dt}\Big)
\end{equation*}

\vspace{7pt}

Done.
\label{pop}
\end{example}

\noindent\rule{\textwidth}{3pt}

\subsection{Analysis}

To facilitate analysis, the nonlinear kernel for function $h(s,t)$ will be evaluated with definite limits $\{0,t\}$ and the constant of integration $C(s)$ set to unity, \emph{viz.}:
\begin{equation}\label{specificsoln}
\begin{split}
 h(s,t)= \Big(1-(p-1)\epsilon\, &  \int_0^t{g^{(p-1)}e^{-b(p-1)t}dt}\Big)\\
 = \Bigg(1-\frac{\epsilon\,  }{b+ D
   (2 \pi s)^2}+\epsilon\, &\frac{e^{-(p-1)\left(D  (2 \pi s)^2 t+bt\right) }  }{b+D (2 \pi s)^2}\Bigg),
 \end{split}
\end{equation}
\noindent the reciprocal integrating factor: $e^{(p-1)bt}$, has been pulled out of the definition for convenience.

Even though the definition for the solution, function $h(s,t)$, is completely arbitrary, this definition holds some nice properties. Firstly, in the limit of time to zero, the function $h(s,t)$ equals unity, which yields a Dirac delta function, $\delta(x)$, after inverting from the Fourier codomain to the original spatial domain. This means the initial condition is set by some arbitrary magnitude $A$, i.e. $u(x,0)=A\delta(x)$. Furthermore, this also means the impulse solution for Green's heat function is valid for this nonlinear kernel solution, therefore, the kernel solution constitutes a nonlinear Green's function or Green's solution. Further complications are thereby possible by employing the transfer function theorem with theorem (\ref{theoremconvsoln}). 

As a means of checking if the solution reduces to the linear case, setting the nonlinear medium coefficient $\epsilon$ to equal zero, the solution in theorem  (\ref{theoremconvsoln}) immediately reduces to the linear solution with convection, \emph{viz.}:
\begin{equation*}
 u(x,t)=G(x,t)e^{-bt}; \{\epsilon=0\}
\end{equation*}

If the medium response is solely linear, then it would be expected both the equation and the general solution would reduce to the linear case\ldots as demonstrated. 

With that said, let's see the behavior of the nonlinear solution along the boundary of the domain, where special attention is given to see whether or not the maximum principle is satisfied. In the limit of infinite time, the general solution, equation (\ref{gensol1}), would approach the steady state. By inspection, the specific solution, eq. (\ref{specificsoln}), shows the function $h(s,t)$ approaches a constant in time, in the codomain, this constant is multiplied by Green's function and the exponential dependent on the variable $b$, moreover, the frequency response of the function $u(s,t)$ is dominated by Green's function in the codomain, therefore, so the same in the original spatial domain; as a consequence, the entire solution approaches zero along the boundary of the domain, $\partial\mathcal{D}$. The solution, even though it represents a nonlinear equation, still satisfies the maximum principle, because Green's function dominates the solution for large time. It is imperative for the heat equation, regardless if it is linear or nonlinear in nature, to satisfy the maximum principle, because for both elliptic and parabolic partial differential systems, harmonic theory states for any precompact subset $\mathcal{D}$ of the domain $u(x,t)$, the maximum of $u(x,t)$ on the closure of $\mathcal{D}$ is achieved on the boundary of $\mathcal{D}$, i.e. $\partial\mathcal{D}$.  

The spatial behavior can be analyzed in the codomain by focusing on Green's kernel in the nonlinear kernel, eq. (\ref{specificsoln}), which generally sharpens as the power of non\textendash linearity, \textit{p}, increases; this indicates the energy is being swiftly spread along the spatial dimension. This behavior is to be expected, given the original differential system, eq. (\ref{diffeqconv}), is a repeated convolution, hence, the solution $u(x,t)$ is continuously being folded in upon itself over time, that is to say, the memory of the system cause the spectrum to widen with increasing power of non\textendash linearity. 

There exists a very interesting behavior with respect to the magnitude of the nonlinear medium response, $\epsilon$. The ramifications of this behavior have greater meaning for hyperbolic differential systems, which parabolic partial differential systems are a basis for such higher order partial differential systems. Nonetheless, with respect to the system discussed in this monograph, the magnitude of non\textendash linearity results in a curious behavior for certain values for the medium coefficient $\epsilon$. If we solve for the root of equation (\ref{specificsoln}), we find the following general condition, with frequency variable set to zero, \emph{viz.}:
\begin{equation}\label{root}
 t_0=\frac{\log \left(\mid\frac{\epsilon}{\epsilon-b}\mid\right)}{b p-b};\{\epsilon\in\mathbb{R}\},\{b|b\in\mathbb{R}\, \&\, b\geq 0\},\{p|p\in\mathbb{Z}\, \&\, \mathbb{Z}>1\}
\end{equation}

The condition above gives the time value, $t_0$, for when the function $h(s,t)$ will cross the origin along the time axis as a function of variables $\{\epsilon,p,b,C(s)=1\}$, and, since the function $h(s,t)$ is really the denominator in the solution, a root creates an asymptotic discontinuity in time for critical times $t_0$. By inspection of definition, eq. (\ref{specificsoln}), if $\epsilon$ is negative, the denominator will always be positive, thus there are no roots; but, for positive values, especially larger than unity, specifically,  $\epsilon>C(s)$, the function $h(s,t)$, starting from the time origin, the function decreases from its maximum, $C(s)$, crossing the time axis at time $t_0$, thereafter, becoming a strictly negative function of time. If a root exists, the value of function $h(s,t)$ will be positive as it approaches the root from the left and negative as it continues beyond the root. This will cause a momentary explosion in the magnitude for $u(x,t_0)$; but, this odd behavior will only occur for time, $t_0$, contained in the domain, $\{t_0|t_0\in [0,\infty)\}$. The time for $t_0$ is either zero or negative for typical values $\{0\leq|\epsilon|\leq 1\}$, and, by inspection, this eliminates any root occurring within the domain, although, the convection constant can modify these results. 

The heat equation is generally a dissipative system, therefore, for all negative values of $\epsilon$, the root, $t_0$, is negative and exhibits no roots; thus, for all regular definitions for the medium response, regardless if it is linear or nonlinear, the solution is regular, convex, monotonic and bounded on the domain $D$. For values of $\epsilon>C(s)$, there does exists a root, $t_0$, within the time domain, hence, there will be an asymptote somewhere along the timeline. This odd behavior is truly a nonlinear resonance and has specific application for certain physical problems.

Scientists often employ the original equation, eq. (\ref{diffeqconv}), with coefficient $\epsilon\ll 1$ as a model in biological sciences, where two such examples are 1) genetic drift (alleles or other biomarkers) shifting in a population over time, modeling the invasion of a new dominate allele throughout a given population of some species of animal, and, 2) the shift in prevalence for some species of animals in a population over some environmental area, given food sources, disease, species invasion and other extraneous factors. The existence of a momentary explosion in population is certainly interesting and warranted, for many real\textendash world problems show sudden blooming of populations, whether it be viruses, bacteria or other larger animal species considered. 

Equation (\ref{root}) provides the root along the timeline for values $\{b,\, \epsilon\}$. In the limit of $\epsilon$ to unity from the righthand side, the root in time approaches infinity; thus, the effect of any asymptotic root would not be visible. The overall affect of the nonlinear medium response is greatly muted for roots downfield along the timeline, because Green's function diminishes rapidly and essentially overtakes any awkward nonlinear influence. For all values larger than unity, the root swiftly moves to the origin, the root causes an asymptote located at time $t_0$ specified by eq. (\ref{specificsoln}). Because of the asymptotic behavior occurring near the origin, the Green's function still yields reasonable values, therefore, the result can cause a momentary, explosive increase along the asymptotic line in time. Typically, biologists use eq. (\ref{diffeqconv}) with values of $|\epsilon|<1$, which prove solutions that are smooth, regular and monotonic; yet, even slightly larger values produce muted nonlinear responses; the greatest affects regarding non\textendash linearity are seen for values $\{C(s)=b=1,\epsilon=2\}$ and $p=2$. 

Generally speaking, the solution $u(x,t)$ behaves similar to the linear system, but the nonlinear kernel tends to cause the impulse to spread spatially much faster than the linear system and also vanish in time much faster than the linear system would. The larger the order of non\textendash linearity, $p$, the faster the system dampens to zero, hence, the energy of the system is swiftly dissipated in short order. In fact, in the limit of infinite order of non\textendash linearity, the nonlinear function $h(s,t)$ approaches unity, \emph{viz.}:
\begin{equation*}
 \lim_{p\rightarrow\infty}\left(\frac{1}{\Big(C(s)-\epsilon\, (p-1)\int{g^{(p-1)}e^{-b(p-1)t}dt}\Big)^{1/(p-1))}}\right)\rightarrow \frac{1}{\left(C(s)\right)^0}\rightarrow 1,
\end{equation*}
\noindent any constant raised to the power of zero yields unity, therefore, the inverse Fourier transform is a Dirac delta function, $\delta(x)$.

This result confirms the influence of nonlinear medium response in thermodynamic systems vanish for large orders of non\textendash linearity, therefore, it is wise to keep this in mind when contemplating such systems, for it is common to erroneously think the larger the order of non\textendash linearity, the larger the influence or the larger the distortion from the linear system, in fact, the very opposite has been clearly demonstrated. 

As for the theory of distributions, analyst typically look for large values and find the natural tendency for the distribution toward a normally  distributed set, the \emph{Central Limit Theorem}; yet, the above analysis does show in the extreme, that is to say, for an infinite convolution of some distribution, the limit is unity. This result makes perfect sense, given any distribution, the infinite iterated convolution would lead to an ever widening variance.

\subsection{Further complications}
If further complication of the general solution be sought, there are two important perspectives to consider. The first is the use of the general solution as an impulse response (\emph{kernel}) to represent non\textendash linear systems in a transfer function. 

Traditionally, if a forcing function is involved, that is to say, if the differential system be considered inhomogeneous, the manner in which the influence of some arbitrary forcing function $K(x,t)$ is handled is by integrating the forcing function by the complementary solution to the heat equation $u(x,-t)$, then multiplying that integral result by the solution $u(x,t)$, for example, in the linear case, \emph{viz.}:
\begin{align}\label{lineq}
\frac{\partial}{\partial t}u(x,t)-& D\frac{\partial^2}{\partial x^2}u(x,t)+b\, u(x,t)=K(x,t),\\\label{genlinsoln}
 u(s,t)= & 
e^{-D(2\pi s)^2t}e^{-bt}\int{k(s,t)e^{D(2\pi s)^2t}e^{bt}dt}+B(s)e^{-D(2\pi s)^2t}e^{-bt},\\
K(x,t)\subset\; & k(s,t);\; u(s,0)=B(s)
 \end{align}

The forced linear heat equation, eq. (\ref{lineq}), has general solution, eq. (\ref{genlinsoln}), where the second term describes how the initial condition, that is to say, the initial distribution of heat throughout space at time zero, dissipates in time and space, and, is realized by convolving $B(x)$ with the heat kernel. The first term in eq. (\ref{genlinsoln}) is the influence of the forcing function, proper, where the forcing function $k(s,t)$ is integrated over time with the complementary solution to the linear heat equation, finally, this integral is convolved by the linear heat kernel to represent how the forcing function dissipates over time and space. It is imperative to understand the meaning of the maximum principle, which states there is only one maximum within the domain for the heat equation. Consider the forcing function being active for some window of time, say, $0\le t\le t_0$, the particular solution, the first term in equation (\ref{genlinsoln}), would be active up until time $t_0$ and thereafter would be zero; thus, in applying any time dependent forcing function, it is advised to keep in mind the last profile forced becomes the initial distribution, $B(x)$, for the second term in equation (\ref{genlinsoln}), which would then describe how the forced distribution of heat will dissipate over time.

In the case of the nonlinear system contemplated, eq. (\ref{diffeqconv}), a very similar definition ensues, as was seen in the linear case; but, the forms involved are decidedly more complex, \emph{viz.}:. 
\begin{equation*}
\begin{split}
\frac{\partial}{\partial t}u(x,t)-& D\frac{\partial^2}{\partial x^2}u(x,t)+b\, u(x,t)-\epsilon\, \Big(u(x,t)\ast_p\Big)=K(x,t),\\
& \\
 u(s,t)= & \, 
e^{-D(2\pi s)^2t}e^{-bt}\Big(h(s,t)\Big)^{-1/(p-1)}\int{k(s,t)e^{D(2\pi s)^2t}e^{bt}\Big(h(s,-t)\Big)^{-1/(p-1)}dt}+\\
& \vspace{-7pt}\\
& \hspace{2in} B(s)e^{-D(2\pi s)^2t}e^{-bt}\Big(h(s,t)\Big)^{-1/(p-1)},\\
K(x,t)\subset\, & k(s,t);\; u(s,0)=B(s)
 \end{split}
\end{equation*}

\noindent\rule{.25\textwidth}{1pt}

\newpage

Despite the complexity, the benefit of achieving closed\textendash form, analytic solutions can not be understated: such results provide the analysts far greater insight into the overall behavior for systems under study; additionally, analytic results provide far greater control over any subsequent integration, where efforts can be mounted to minimize numerical error; lastly, the ability to provide Pareto or better with respect to the total energy or information represented by the original partial differential system can be priceless. Achieving Pareto, eighty percent or more of the total energy or information represented by a system, enables tremendous advantages, where pure numerical integration, whether finite difference methods, finite element or other some other purely numerical method are known to be costly in time, resource hungry and fraught with frustrations; such methods never are able to guarantee the numerical results obtained are accurate, worse, analysts are often without any closed\textendash form, analytic solution, to wit, a comparison may be made, therefore, are flying completely blind.

Additional complications, imaginable, would be an unknown function, \emph{per chance}, comprised of a set of functions, i.e.
\begin{equation*}
 u(x,t)=\Big(f(x,t)\ast G(x,t)\ast K(x,t)\Big)(x),
\end{equation*}
\noindent where function $K(x,t)$ is arbitrary. 

Three main scenarios arise in considering an additional set of functions to the substitution. The first would be a time dependent function $K(x,t)$. The condition of canceling the second order spatial derivative would still be satisfied, but the nature of function $K(x,t)$ would need be known in order to work out in detail the reduced differential system. Ultimately, the Bernoulli differential equation would result and the specifics would need be worked by the reader who would contemplate such complexity.

The second general scenario would be the function $K(x)$ solely dependent on the spatial variable. There are two general cases one might consider where function $K(x)$ is a set of functions either convolved or multiplied, \emph{viz.}:
\begin{align*}
  K^{(1)}(x)=&\Big(K_1K_2\ldots K_n\Big)\\
  K^{(2)}(x)=&\Big(K_1\ast K_2\ast\cdots\ast K_n\Big)
 \end{align*}

Both substitutions would still satisfy the requirement for canceling the second order spatial derivative. The solutions would be the following two definitions:
 \begin{align*}
 & u(x,t)=\, \Big(G(x,t)\ast K^{(1)}(x)\ast\mathscr{F}^{-1}\Big\{F(s,t)\Big\}\Big),\, F(s,t)=h^m,\,\, m=-1/(p-1),\\
  & K^{(1)}(x)=\, \Big(K_1K_2\ldots K_n\Big)\subset\Big(\hat{k}_1\ast\hat{k}_2\ast\cdots\ast\hat{k}_n\Big),\\
  & h(s,t)=\, e^{b(p-1)t}\Big(C(s)-\epsilon\, (p-1)\int{\Big(\hat{k}\ast_{(n-1)}\Big)^{(p-1)}g^{(p-1)}e^{-b(p-1)t}dt}\Big),
 \end{align*}
 \noindent where $\hat{k}$ represents the Fourier transform of the Kernel function considered.

 In the second case, a bit of slight of hand will be employed to show the reader explicitly these nonlinear solutions are not truly dependent on the substitution, \emph{per se}, rather, the nonlinear kernel is a simple exponential mapping to represent the nonlinear term, whose coefficient is $\epsilon$, \emph{viz.}:
 \begin{align*}
 & \frac{\partial}{\partial t}u(x,t)- D\frac{\partial^2}{\partial x^2}u(x,t)+b\, u(x,t)=\epsilon\, \Big(u(x,t)\ast_p\Big)\ast\Big(K_1\ast K_2\ast\cdots\ast K_n\Big),\\
 & u(x,t)=\, G(x,t)\ast\mathscr{F}^{-1}\Big\{F(s,t)\Big\},\\
 & F(s,t)=h^m,\, m=-1/(p-1),\\
  & K^{(2)}(x)=\, \Big(K_1\ast K_2\ast\cdots\ast K_n\Big)\subset\Big(\hat{k}_1\hat{k}_2\ldots\hat{k}_n\Big)=\hat{k},\\
  & h(s,t)=\, e^{b(p-1)t}\Big(C(s)-\epsilon\, (p-1)\int{\hat{k}^{(n)}g^{(p-1)}e^{-b(p-1)t}dt}\Big)
 \end{align*}

This last manipulation takes into account the nonlinear exponential mapping is wholly dependent upon the definition for the nonlinear term alone, hence, one may think of this as normalizing the solution $u(x,t)$ by dividing by the additional function $K^{(2)}(x)$. The above manipulation shows the versatility of the general solution provided and is only limited by the imagination of the analyst. If serious thought be given, the solution to the nonlinear equation really is comprised of a nonlinear kernel affected by only the expressed definition for the nonlinear term, therefore, considerable freedom exists in just manipulating the solution to satisfy additional differential systems. 
\vspace{5pt}
 
The set of functions $K^{(1,2)}(x,t)$ could be Bernoulli trials, Poisson distributions, gamma distributions, \&c, or any combination thereof; of course, whether this combination would be amenable to integration and analytic representation is wholly dependent on the definitions and the resulting integrand, plus, many other factors.

\subsection{An illustration}
As an illustration and as a general aid, the general solution will be applied for a Fisher's type equation. With nonlinear power $p$ equal to two, the convolution equation generates a nonlinear medium response represented by the self convolution of the solution; assuming parameters $\{D,b,\epsilon\}$ are all positive, the following definition results:
 \begin{equation}\label{fisherc}
 \frac{\partial}{\partial t}u- D\frac{\partial^2}{\partial x^2}u+b\, u-\epsilon\, \Big(u\ast u\Big)=0 
\end{equation}

Referring to the general solution, theorem (\ref{theoremconvsoln}), function $h(s,t)$ takes on the following form after substituting the requisite parameter values: 
\begin{equation}\label{fisherceq}
 h(s,t)= \quad\frac{e^{-bt}}{1+\epsilon\, \int{g\, e^{-bt}dt}\Big\vert_{t=0}-\epsilon\, \int{g\, e^{-bt}dt}}
 \end{equation}

In general, the inverse Fourier transform of this formula is unknown; but, most researchers study systems where the leading nonlinear coefficient, $\epsilon$, is much less than unity, therefore, the Binomial theorem will be employed to expand equation (\ref{fisherceq}).

The Binomial theorem is stated thus:
\begin{equation*}
 \frac{1}{(x+y)^n}=\sum_{k=0}^{\infty}\binom{-n}{k}x^{-n-k}y^{k}=\sum_{k=0}^{\infty}(-1)^k\binom{n}{k}x^{n-k}y^{k}
\end{equation*} 

If we consider the second and third terms in the denominator of equation (\ref{fisherceq}) both to represent parameter $y$ in the Binomial expansion, then we may justify an expansion under arguments: both terms cancel one another for time to zero; for larger times, the third term vanishes; the second term has its largest value moderated by the nonlinear coefficient, thus, $\epsilon/b$; lastly, since the nonlinear coefficient is much less than unity, $\epsilon\ll 1$, this term is guaranteed smaller than unity (\emph{assuming $b=1$}). Also, the limits of integration will be assumed to be $\{t|0\le t\le t\}$. 

The expansion yields two terms, plus, higher order terms, which will be neglected, since the series is a Cauchy series, \emph{viz.}:
\begin{equation*}
 h(s,t)\approx\; 1+\frac{\epsilon}{b+D(2\pi s)^2} -\epsilon\, \frac{e^{-D(2\pi s)^2 t-bt}}{b+D(2\pi s)^2}+\, \textit{higher order terms},
\end{equation*}
\noindent where the exponential, $\exp\left(-bt\right)$, will be considered momentarily. Multiplying by the Gaussian function and the exponential, yields:
\begin{equation*}
 u(s,t)\approx\; e^{-D(2\pi s)^2t}e^{-bt}+\epsilon\, \frac{e^{-D(2\pi s)^2t}e^{-bt}}{b+D(2\pi s)^2} -\epsilon\, \frac{e^{-2D(2\pi s)^2t}e^{-2bt}}{b+D(2\pi s)^2}
\end{equation*}

This expansion enables ease in finding the inverse Fourier transform for each term, \emph{see ''Formulae'' Table in Appendix (\ref{Appendix})}. The inverse Fourier transform of the solution yields the following equation, \emph{viz.}:
\begin{equation*}
\begin{split}
 u(x,t)\approx\; & G(x,t)e^{-bt}+\epsilon\, \frac{e^{-\left(x\sqrt{b/D}\right)}}{4\sqrt{Db}}\, \mathrm{\textit{erfc}}\left(\frac{2\sqrt{Db}\, t-x}{2\sqrt{Dt}}\right)+ \\
  &  \hspace{2in} \epsilon\, \frac{
 e^{\left(x\sqrt{b/D}\right)}}{4\sqrt{Db}}\, \mathrm{\textit{erfc}}\left(\frac{2\sqrt{Db}\, t+x}{2\sqrt{Dt}}\right)-\\
 & \epsilon\, \frac{e^{-bt}e^{-\left(x\sqrt{b/2D}\right)}}{4\sqrt{2Db}}\, \mathrm{\textit{erfc}}\left(\frac{2\sqrt{2Db}\, t-x}{2\sqrt{2Dt}}\right)- \\
  &  \hspace{2in} \epsilon\, \frac{
 e^{-bt}e^{\left(x\sqrt{b/2D}\right)}}{4\sqrt{2Db}}\, \mathrm{\textit{erfc}}\left(\frac{2\sqrt{2Db}\, t+x}{2\sqrt{2Dt}}\right)/;(\epsilon\ll 1)
 \end{split}
\end{equation*}

By inspection, the solution reduces to the linear solution for $\epsilon$ equal to zero\textemdash continued validation is imperative. The limit for the complementary error function is zero, irrespective of the limit for either the spatial or time variable to infinity, thus, the solution approaches zero along the boundary of the domain $\partial\mathscr{D}$. In the limit of zero time, the second and third terms cancel, plus, Green's function, $G(x,t)$, approaches a Dirac delta function of spatial variable \textit{x}, i.e. $\delta(x)$. It must be remembered, this solution represents the impulse drive for the nonlinear system, thus, further complications are possible through additional convolutions.

As time expands from zero, the third term will decline in value, increasing the reduction in magnitude by the second term. In the spatial axis, the exponential raised to the power of the spatial variable will expand the energy along the x-axis, hence, showing the tendency for convolutional systems to spread the energy swiftly over the spatial dimension. Since the error function is the integral of the Gaussian function, it has the property of rising from its floor value to its maximum value of unity rapidly, therefore, at some time, $t_0$, the nonlinear system swiftly shutoffs and becomes quiescent.

\section{The Classic Newell\textendash Whitehead\textendash Segel Equation}
The classic form for the Newell\textendash Whitehead\textendash Segel equation has nonlinear medium response expressed in multiplicative form, \emph{viz.}:
\begin{equation}\label{classic}
 \frac{\partial}{\partial t}u- D\frac{\partial^2}{\partial x^2}u+b\, u-\epsilon\, u^p=0, 
\end{equation}
\noindent where $p$ is a non\textendash negative integer greater than unity, i.e. $\{ p | p \in \mathbf{Z}^{\neq} \text{ and } \mathbf{Z} >1  \}$. 

The general solution for this equation causes considerable consternation. The equation does not allow separation. If a product of two functions, both a function of the dependent variables, be substituted for the unknown function $u(x,t)$, the very necessary simplification that removes the second order spatial derivative is not typically possible. In fact, a product substitution makes matters worse, where four terms arise containing first and second order derivatives, \emph{viz.}:
\begin{equation*}
 \frac{\partial^2}{\partial x^2}\left(Gf\right)=\left(G_{xx}f\right)+\left(Gf_{xx}\right)+\left(G_x f\right)+\left(Gf_x \right),
\end{equation*}
\noindent assuming $u(x,t)=G(x,t)f(x,t)$, and, the subscript signifies the order of derivative.

The above expansion, in and of itself, is not necessarily the greatest concern; it turns out, multiplication of the unknown function $u^p$ is what causes the greatest trouble in attempting a solution. Multiplication transforms to a set of repeated convolutions in the codomain, whose resulting $p^{\text{th}}$ convolution is generally not known. Now, in the case of a Gaussian function, \emph{inept}, the Fourier transform of Green's function, repeated convolutions can be calculated, because the function is convex and bounded on the open interval.  

The best path to a general solution, as seen for the iterated convolution case, must transform the original differential system into the Bernoulli differential equation. There are many approaches by which one may transform the original equation; but, typically speaking, it is advisable to cast the Bernoulli differential equation in a form of products, one known and the other yet defined. As was seen for the $p^{\text{th}}$ convolutional differential system, the Fourier transform, transformed the convolutions into a product of functions, which enabled ready solution of the problem. In the present system, the multiplicative form, the Fourier transformation will transform the product into a set of iterated convolutions. This bodes badly for solving the resulting Bernoulli differential equation. As a consequence, serious consideration must be given beforehand as to the form of the functions employed. Ultimately, it is highly desired the $p^{\text{th}}$ convolution should result in a simple function of some sort, thereby, fascilitating ease in further complications. 

With this end in mind, consider the folowing $n^{\text{th}}$ root form of Green's function, \emph{viz.}:
\begin{equation*}
  G'=\left(\frac{e^{-\frac{x^2}{4 D\, t}}}{\sqrt{4\pi D\, t}}\right)^{\left(\frac{1}{n}\right)}\subset \frac{\sqrt{n}\, e^{-D(2 \pi s)^2 n\, t}}{\left(4\pi D\, t\right)^{\frac{1-n}{2n}}}=g',
\end{equation*}
\noindent where a prime is attached to both the function itself and its Fourier transform, indicating this special form of Green's function. 

At first sight, this form of Green's function may appear completely unhelpful, but it holds a very special property, namely, convolving the Fourier transformed form $n-1$ repeated times yields the regular transformed Green's function, \emph{viz.}:
\begin{equation*}
 \Big(g'\ast_{(n-1)}\Big)=g\supset G(x,t)
\end{equation*}

It is with the foresight of knowing the multiplication in equation (\ref{classic}) will transform to yield a $p^{\text{th}}$ iterated convolution, that the $n^{\text{th}}$ rooted form of Green's function will be employed, thereby, transforming the original equation into the Bernoulli differential equation.   

To start, the substitution for the unknown solution $u(x,t)$ will be a product of a function of time and Green's rooted function, i.e.
\begin{equation*}
 u(x,t)=f(t)\left(\frac{e^{-\frac{x^2}{4 D\, t}}}{\sqrt{4\pi D\, t}}\right)^{\left(\frac{1}{n}\right)}
\end{equation*}

The root will participate in the derivatives, hence, for ease, equation (\ref{classic}) is redefined thus:
\begin{equation*}
 \frac{1}{n}\frac{\partial}{\partial t}u-D\frac{\partial^2}{\partial x^2}u+bu-\epsilon\, u^p=0
\end{equation*}

Taking the equation to the Fourier codomain, yields:
\begin{equation*}
 \frac{1}{n}\frac{\partial}{\partial t}u+D(2\pi s)^2u+bu-\epsilon\, \Big(u\ast_p\Big)=0,
\end{equation*}
\noindent whereupon substitution, yields the following:
\begin{equation*}
\begin{split}
 \frac{1}{n}f_t(t)g'-\frac{1}{2tn^2}f(t)g'+ &\frac{1}{2tn}f(t)g' -D(2\pi s)^2f(t)g'+D(2\pi s)^2 f(t)g'+\\
    & bf(t)g'-\epsilon\, f(t)^p\Big(g'\ast_p\Big)=0,
    \end{split}
\end{equation*}
\noindent where subscript \textit{t} indicates a derivative with respect to time, i.e. $f_t(t)$.

The first four terms are generated by taking the time derivative of the substitution with respect to the time variable, and the last term of the four cancels with the Fourier transformed term involving the second order spatial derivative. Placing the fraction $1/n$ aids in normalizing the fourth term, in order to, facilitate cancellation of the fifth term\ldots the best that can be accomplished. This yields the following reduced equation to solve:
\begin{equation*}
 \frac{1}{n}f_t(t)g'+\left(b+\frac{1}{2tn}-\frac{1}{2tn^2}\right) f(t)g'=\epsilon\, f(t)^p\Big(g'\ast_p\Big)
\end{equation*}

Now, focusing on the repeated convolution, preceded by coefficient $\epsilon$, if the value of $p$ be equal to $n-1$, then the repeated convolution results in the standard Fourier transform of Green's function, \emph{viz.}:
\begin{equation*}
 \frac{1}{n}f_tg'+\left(b+\frac{1}{2tn}-\frac{1}{2tn^2}\right) fg'=\epsilon\, f^pg
\end{equation*}

Dividing through by the rooted Green's function, $g'$, produces a difference of exponents, plus, some additional constants and time variables, thus:

\begin{equation*}
 \frac{g}{g'}=\frac{\left(
 4\pi D\, t\right)^{
 \frac{1-n}{2n}}}{\sqrt{n}}
 e^{-D(2\pi s)^2t}
 e^{D(2 \pi s)^2 n\, t}=\frac{\left(
 4\pi D\, t\right)^{
 \frac{1-n}{2n}}}{\sqrt{n}}
 e^{-(1-n)D(2\pi s)^2t}
\end{equation*}

and multiplying through by $n$ finally yields:

\begin{equation*}
 f_t+\left(n\, b+\frac{1}{2t}-\frac{1}{2tn}\right) f=\epsilon\, \sqrt{n}\left(
 4\pi D\, t\right)^{
 \frac{1-n}{2n}}
 f^pe^{-(1-n)D(2\pi s)^2t}
\end{equation*}
%\begin{equation*}
 %f_t+\left(bn+\frac{1}{2t}-\frac{1}{2tn}\right) f=\epsilon\, f^pg^pg'^{-1}
%\end{equation*}

Once again, we need to employ the transform that enables reducing the Bernoulli differential equation to a linear ordinary differential equation. The transformation is achieved by letting $f(t)=h(t)^m$, therefore, $f_t=mh^{m-1}h'$, where prime indicates a derivative with respect to time (\emph{note: all references to the rooted Green's function $g'$ have been eliminated in the final definition of the Bernoulli differential equation, thus prime should not cause any misunderstanding}). Substituting the transformation yields:
\begin{equation*}
 mh^{m-1}h'+\left(n\, b+\frac{1}{2t}-\frac{1}{2tn}\right) h^m=\epsilon\, \sqrt{n}\left(
 4\pi D\, t\right)^{
 \frac{1-n}{2n}}
 h^{mp}e^{-(1-n)D(2\pi s)^2t}
\end{equation*}

Dividing through by $h^{m-1}$ gives the following:
\begin{equation*}
 m h'+\left(n\, b+\frac{1}{2t}-\frac{1}{2tn}\right) h=\epsilon\, \sqrt{n}\left(
 4\pi D\, t\right)^{
 \frac{1-n}{2n}}
 h^{m(p-1)+1}e^{-(1-n)D(2\pi s)^2t},
\end{equation*}
\noindent and, after solving for the condition $m(p-1)+1=0$, the parameter $m=1/(1-p)$.

Using the parameter $m=1/(1-p)$ removes the last occurrence of the function $h$ and leaves the following linear ordinary differential equation to solve, after dividing through by the parameter \textit{m}, \emph{viz.}:
\begin{equation*}
 h'+(1-p)\left(n\, b+\frac{1}{2t}-\frac{1}{2tn}\right) h=\epsilon\, (1-p)\sqrt{n}\left(
 4\pi D\, t\right)^{
 \frac{1-n}{2n}}
 e^{-(1-n)D(2\pi s)^2t},
\end{equation*}

Solving for the integrating factor yields:
\begin{equation*}
 \begin{split}
 & \int{\frac{\partial}{\partial t}h(t)\exp{\Bigg((1-p)\left(n\, b\, t+\frac{ln(t)}{2}-\frac{ln(t)}{2n}\right)}\Bigg)\, dt}=\\
 & \hspace{1in} h(t)\exp{\Bigg((1-p)\left(n\, b\, t+\frac{ln(t)}{2}-\frac{ln(t)}{2n}\right)}\Bigg) 
 \end{split}
\end{equation*}
\noindent where the natural logarithm is involved, i.e. $ln(t)$.

With the integrating factor in hand, it is a matter of multiplying through and integrating over the time variable, \emph{viz.}: 
\begin{equation*}
 h(s,t)=\bigintsss_0^t{\epsilon\, (1-p)\sqrt{n}\left(
 4\pi D\, t\right)^{
 \frac{1-n}{2n}}
 e^{-(1-n)D(2\pi s)^2t}\exp{\Bigg((1-p)\left(n\, b\, t+\frac{ln(t)}{2}-\frac{ln(t)}{2n}\right)}\Bigg)dt}
\end{equation*}

As it stands now, the formula has a mixture of both integers $p$ and $n$, but it would be advantageous to transform to one single integer, and, since the power of non-linearity, $p$, is more relevant for our purposes, then with $n=p+1$, the integral becomes:
\begin{equation*}
 h(s,t)=\epsilon\, (1-p)\sqrt{p+1}\left(
 4\pi D\right)^{
 -\frac{p}{2p+2}}
\bigintsss_0^t{
 \exp\Big(pD(2\pi s)^2t+(1-p^2)bt\Big)t^{-\frac{1}{2}\frac{p^2}{p+1}}dt}
\end{equation*}

The integrand involves an exponential in time, multiplied by the time variable raised to a power; this type of integral is referred to as the exponential integral, which can be cast into the form of an incomplete gamma function. Most critical to note: the exponents for the time variable will be fractional in nature, which avoids poles for the gamma function. Finally, the integral has been multiplied by the reciprocal integrating factor to yield the final answer, \emph{viz.}:

\begin{equation*}
\begin{split}
 &h(s,t)=e^{-(1-p^2)bt}t^{\frac{p^2-p}{2p+2}}\Bigg(\epsilon\, (1-p)\sqrt{p+1}\left(
 4\pi D\right)^{
 -\frac{p}{2p+2}}\left(-1\right)^{(p/3)}\left(b (1-p^2)+ D p (2\pi s)^2\right)^{(\frac{p}{3}-1)}\times \\
 &\hspace{100pt}\Gamma \left(1-\frac{p}{3},\left(b
   (1-p^2)-Dp(2\pi s)^2\right) t \right)\Bigg\vert_0^\infty+C(s)\Bigg),
   \end{split}
\end{equation*}
\noindent where $\Gamma(n,x)$ represents the incomplete gamma function.

\begin{example}[\textbf{General solution for the Newell\textendash Whitehead\textendash Segel equation}]\label{Solnmulti}

A most general solution has thus been obtained, representing a closed\textendash form, analytic solution for a general homogeneous nonlinear parabolic differential equation, whose nonlinear medium response is expressed as a p\textendash times, multiplicative response, viz.:
\begin{equation*}
 \frac{\partial}{\partial t}u- D'\frac{\partial^2}{\partial x^2}u+b'\, u-\epsilon'\, u^p=0, 
\end{equation*}
\noindent where all primed coefficients are multiplied by $\sqrt{p+1}$.

The solution is comprised of a rooted Green's function and a nonlinear kernel function $f(x,t)$, whose inverse Fourier transform is defined to be the inverse of the function $h(s,t)$ raised to the power of parameter $m=1/(1-p)$, \emph{viz.}:
\begin{equation}\label{classicsoln}
 u(x,t)= G(x,t)^{\frac{1}{p+1}}\ast f(x,t)= G^{\frac{1}{p+1}}\ast\mathscr{F}^{-1}\Bigg\{h^{1/(1-p)}\Bigg\},
\end{equation}
\noindent where the inverse Fourier transform is signified by symbol $\mathscr{F}^{-1}$.

Lastly, function $h(s,t)$ is defined as such:
\begin{equation*}
\begin{split}
 & h(s,t)=e^{-(1-p^2)bt}t^{\frac{p^2-p}{2p+2}}\Bigg(\epsilon\, (1-p)\sqrt{p+1}\left(
 4\pi D\right)^{
 -\frac{p}{2p+2}}\times\\
 &\hspace{25pt}\left(-1\right)^{(p/3)}\left(b (1-p^2)+ D p (2\pi s)^2\right)^{(p/3-1)}\times\\
 &\hspace{50pt}\Gamma \left(1-\frac{p}{3},\left(b
   (1-p^2)-Dp(2\pi s)^2\right) t \right)\Bigg\vert_0^\infty+C(s)\Bigg)\\
   &\hspace{3in}/;(p\in\mathbb{Z}^{\neq}\wedge p>1),
   \end{split}
\end{equation*}
\noindent where $\Gamma(n,x)$ represents the incomplete gamma function; $C(s)$ is the coefficient of integration.

More generally, the function $h(s,t)$ is equal to\ldots
\begin{equation*}
\begin{split}
 h(s,t)= & \epsilon\, (1-p)\sqrt{p+1}\left(
 4\pi D\right)^{
 -\frac{p}{2p+2}}e^{-(1-p^2)bt}t^{\frac{p^2-p}{2p+2}}\times\\
 & \hspace{75pt}\bigintsss{
 \exp\Big(pD(2\pi s)^2t+(1-p^2)bt\Big)t^{-\frac{1}{2}\frac{p^2}{p+1}}\, dt}
 \end{split}
\end{equation*}

\vspace{7pt}

Done.
\label{pop3}
\end{example}

$\ast$\xrfill[.5ex]{.4pt}\qedsymbol

\vspace{25pt}
The manner in how the solution was derived, namely, employing the rooted Green's function, resulted in integrands of one form. If Green's function is used in the initial substitution, then the $i^{\text{th}}$ convolution can be calculated to be the following:
\begin{equation*}
 \Big(g\ast_{i}\Big)=\frac{\sqrt{4\pi Dt}\, e^{-\frac{(2\pi s)^2Dt}{i+1}}}{(4\pi D t)^{(i+1)}\sqrt{i+1}},\, i=\{1,2,3\ldots\}
\end{equation*}

As a consequence, the solution will lead to an integrand comprised of another form. I mention it here for the reader, because there are pros and cons for each definition; nevertheless, the reader can certainly ascertain no simple analytic representation results for the multiplicative form of the differential equation.

\begin{cor}[\textbf{General solution for the Newell\textendash Whitehead\textendash Segel equation}]\label{alter}

A most general solution has thus been obtained, representing a closed\textendash form, analytic solution for a general homogeneous nonlinear parabolic differential equation, whose nonlinear medium response is expressed as a multiplicative response of power $p\in\mathbb{Z}^{\neq}\wedge p\ge 2$, viz.:
\begin{equation*}
 \frac{\partial}{\partial t}u- D\frac{\partial^2}{\partial x^2}u+b\, u-\epsilon\, u^p=0
\end{equation*}

The solution is comprised of Green's function and a nonlinear kernel function $f(x,t)$, whose inverse Fourier transform is defined to be the inverse of the function $h(s,t)$ raised to the power of parameter $m=1/(p-1)$, \emph{viz.}:
\begin{equation*}
 u(x,t)= G(x,t)\ast f(x,t)= G\ast\mathscr{F}^{-1}\Bigg\{h^{1/(1-p)}\Bigg\},
\end{equation*}
\noindent where the inverse Fourier transform is signified by symbol $\mathscr{F}^{-1}$.

Lastly, function $h(s,t)$ is defined as such:
\begin{equation}\label{expsoln}
 h(s,t)=  \epsilon\, (1-p)e^{bt}
 \bigintsss{
 \frac{\sqrt{4\pi Dt}\, e^{-\frac{(2\pi s)^2Dt}{p-1}}}{(4\pi D t)^{p-1}\sqrt{p-1}}e^{-bt}dt}
\end{equation}

\vspace{7pt}

Done.
\label{corrol}
\end{cor}

$\ast$\xrfill[.5ex]{.4pt}\qedsymbol

\vspace{7pt}

The affect of continued convolutions are most easily seen in the Corollary definition for function $h(s,t)$, where the spectrum will widen with rising power $p$, leading to more concentration in the original spatial domain.

Taking the first order series approximation for integer \textit{p} and time to infinity, with respect to the logarithm of the integrand in equation (\ref{expsoln}), yields a function, whose limit for integer \textit{p} to infinity and infinite time is negative infinity, \emph{viz.}:
\begin{equation*}
 \lim_{p\rightarrow\infty}\, \ln{\Bigg(\Big(\bigintsss{
 \frac{\sqrt{4\pi Dt}\, e^{-\frac{(2\pi s)^2Dt}{p-1}}}{(4\pi D t)^{p-1}\sqrt{p-1}}e^{-bt} dt\Big)^{1/
(1-p)}}\Bigg)\Bigg\vert_{\{p,t\}\rightarrow\infty}}\sim\; -\infty
\end{equation*}

The asymptotic expansion tends to negative infinity, therefore, the exponential magnitude is decreasing for large integer \textit{p} and time. The proportionality shows for large time and power of non-linearity, integer $p$, the kernel for the function $h(s,t)$ approaches zero, \emph{viz.}:
\begin{equation*}
 \lim_{\{p,t\}\rightarrow\infty}{\; h(s,t)}^{1/(1-p)}\rightarrow 0
\end{equation*}

There are certainly a range of values for integer \textit{p} and time to be explored; but, in the extreme limit, the solution $u(x,t)$ goes to zero. This is another verification that large orders of non-linearity lead to contradictory results.

\subsection{An illustration}
The classic Fischer's equation will be used for illustration purposes. The equation is primarily employed in the biological sciences to model genetic drift in a population \cite{fisher}. For a set of alleles, whose respective probabilities for dominance are, say, \emph{p, q, r}\ldots; the corresponding partial differential model is defined by R.A. Fisher, \emph{viz.}:
\begin{equation*}
 \frac{\partial}{\partial t}u= D\frac{\partial^2}{\partial x^2}u+\epsilon\, upqr\ldots
\end{equation*}

Assuming the probabilities are a function of space and time, the solution is immediately the time integral, where the integrand includes the convolution of each respective probability for each allele, \emph{viz.}:
\begin{equation*}
 u(x,t)=G(x,t)\ast\mathscr{F}^{-1}\left\{\exp\left(\epsilon\, \int{p(s,t)\ast q(s,t)\ast r(s,t)\ast\ldots dt}\right)\right\}
\end{equation*}

Since Green's function has been employed; then, for time zero, the limit is a Dirac delta function of space. The model was originally intended to model allele propagation through a population in an environment, say, birds along a shoreline; but, since Green's function has been employed, one could consider the model for allele substitution within a genetic code. This would require a spatial dependent probability for each allele, for example, $p(x,t)$; therefore, the probability function would describe the likelihood for substitution or insertion at specific locations along the genetic sequence. 

If the spatial dependence be dropped and the allele probability is time dependent only, then the solution would require integrating over that time dependence. Finally, if we assume the allele probabilities are constants, then the solution simplifies considerably, \emph{viz.}:
\begin{equation*}
 u(x,t)=G(x,t)e^{\epsilon\, pqr\ldots\, t}
\end{equation*}

The solution does not satisfy the maximum principle because the leading coefficient $\epsilon$ is defined to be a positive real number, $\epsilon\in\mathbb{R}^+$. In its simplest form, it is debatable if this model has any meaningful application to physical processes.

One complication that could bring greater meaning to Fisher's equation would be to raise the power of the unknown solution in the last term, thereby, converting the differential equation to a nonlinear partial differential equation, whose solution is known. Consider the following:
\begin{equation*}
 \frac{\partial}{\partial t}u= D\frac{\partial^2}{\partial x^2}u+\epsilon\, u^2pqr\ldots
\end{equation*}

The solution is immediately known, \emph{viz.}:
\begin{equation*}
 \begin{split}
  u(x,t)= & G(x,t)\ast\mathscr{F}^{-1}\left\{h(s,t)^{-1}\right\}\\
  h(s,t)= & \exp\left(\epsilon\, \int{\Big(g\ast\hat{p}\ast\hat{q}\ast\hat{r}\ast\ldots\Big) dt}\right),
 \end{split}
\end{equation*}
\noindent where the hat signifies the Fourier transform, i.e. $\hat{p}=p(s,t)$, also, small \textit{g} symbolizes Green's function in the codomain, i.e. $G\subset g$.

The solution represents the probability of an allele substitution or dominance, with spatial affinity specified over some time window. The intensity for selection, coefficient $\epsilon$, may now be any positive or negative real number, i.e. $\epsilon\in\mathbb{R}$. The overall solution also satisfies the maximum principle, which is essential for any meaningful application to some real\textendash world physical process. It is thoroughly possible to imagine a wide set of possible analytic solutions, given a set of representations for probabilities: \emph{p, q, r}\ldots, including the inverse Fourier transform.

\section{Epilogue}
Computer aided numerical analysis offers many advantages to the young researcher, but new technology can hamper as much as it can aid. The most dogmatic of computer scientists/advocates ardently believe, with enough computer power, Laplace's daemon could be achieved. Virilly, many, both scientific and layman alike, being technologically spellbound, suffer the decisive inability for critical thought in modern times; and, this mental deficiency grows worse with time as the promise of such daemons, such as, artificial intelligence (AI) should grow ever more gargantuan. 

Laplace was once asked about the potential to model a room full of molecules, atoms and other particles. The classic mechanical descriptions were certainly within hand, Newtonian mechanics, but the sheer number of particles would require the mind of God, no less, to simultaneously hold all parameters in real time. 

\begin{quote}
 {\textit{We may regard the present state of the universe as the effect of its past and the cause of its future. An intellect which at a certain moment would know all forces that set nature in motion, and all positions of all items of which nature is composed, if this intellect were also vast enough to submit these data to analysis, it would embrace in a single formula the movements of the greatest bodies of the universe and those of the tiniest atom; for such an intellect nothing would be uncertain and the future just like the past would be present before its eyes.}
 
 \hspace{1.125in} \textendash\, Pierre Simon Laplace, ''A Philosophical Essay on Probabilities''\cite{Laplace}}
\end{quote}

In some sense, this line of reasoning is curious coming from a man known for both his knowledge of probability theory and personal penchant for games of chance. That this great philosopher, Laplace, completely ruled out any possibility randomness could play in the happenings of this universe is more a testimony to the prestige widely attributed to Newtonian mechanics; in those times, no one questioned the preeminence, the rightness, the truth of classical mechanics, not in public at least\textemdash perish the thought! Ironically, randomness has reared its ugly head against those who would claim mastery over the universe, at least, in thought or understanding; but, like a nightmare, randomness proves a more viable, even more real concept of the universe than the more desired deterministic views. Quantum mechanics, of course, destroyed all hopes for a clockwork universe; the universe has forever changed to a big maybe. Frankly, the quantum revolution in scientific thought represents a more mature perception of the world, for determinism is rather juvenile in its nature and desire. 

With that said, these two forms of thought are still at war with one another, where determinism forever seeks dominance. It's a matter of control for those who subject themselves to this type of childish thinking\textemdash determinism\textemdash a type of wishful thinking, a wish for certainty, for a promise\ldots

Modernity has, somewhere along the line, decided to replace human thought with algorithmic thought\textemdash all in an attempt to gain full mastery\ldots but of course. The benefits are seen to far outweigh any possible human contribution; yet, algorithms are designed by humans, thus, suffer bias, cliches and other laughable fits and starts; reminding one of some cantankerous whirligig desperately trying to wind. The very real threat of algorithmic bias demands our human vigilance, because time and bias drift could prove fatal to humanity, if not just embarrassing.

I was once asked to model a particular mercury filter, designed by a certain group in the U.S. Navy. The filter itself was rather crude in construction; but, putting aside the absurd casing of thick PVC plastic piping, the inner filter was an industrial grade fiber filter. Such filters are rated to catch particles of some minimum size, say, micron size. I was tasked with building a model in COMSOL to model mercury laden effluent going into the filter cavity and calculate the ''performance'' of the filter! This was obviously an exercise in pure futility. The answer is decided by the chosen fiber filter and nothing else. It would seem computer\textendash driven reality, that is, virtual reality has taken precedence over the real world that lay around us. What a curious happenstance, indeed!  

As a trained numerical analyst, I too, was once fascinated by the prospect of solving formerly intractable problems through the aid of computers and carefully designed numerical algorithms; contrariwise, my own experiential \emph{corpus} has proven otherwise, inducing me to become a modern Luddite, of sorts. Modern analysts praise the ''all\textendash important'' \underline{data} in their so\textendash called \emph{data\textendash driven policies}, as if points of facts trump all reality; albeit, reality, our human reality, is comprised of data, that is, signals, but a healthy dose of Humean skepticism is required for any worldly, self\textendash respecting man. Data, as it is called or referred to, is no more real than any other signal; in fact, what is referred to as ''data'' is biased by the fact we humans decide it is a ''data point'' in the first place, christening this seemingly vaporous entity as worthy of our attention. It was David Hume who unequivocally proved that man's epistemological limits are small, indeed, and there is no means with which we may reach beyond the threshold, despite all our might. Like the weary widow seeking to contact her lost husband beyond the grave\ldots 

Where David Hume proved all human knowledge imperfect and dependent upon the imperfect senses of the body, it was Friedrich Nietzsche, the German philosopher, who dealt with the epistemological problem in the more modern sense; his analysis was simple: \emph{There is no reality}. Nietzsche speaks of ''facts,'' yet equally demands his readers to ''go beyond Good and Evil,'' that is to say, to set aside their paltry human judgmentalism, for such effete forces are no match for the world that surrounds us\ldots simply put, the forces at play around us are far larger than us. 

History has proved otherwise, for humans are rather tenacious, bullheaded and stubborn, especially, when they seek an object of desire. Determinism, as a philosophy, has never recovered from the revolution of quantum mechanics. The old, comforting Newtonian world, where everything could be known and understood, simply died away in a fortnight\ldots and that left many bereft of any belief. It is \emph{belief} and all its dogmatic undertones at play in this topic, to be sure, and, many zealots cannot accept the death of their messiah; it would be anathema to do so, worse, it would deny the very basis upon which their being finds rest, worse still, the death of their existential wellspring.

An ever encroaching alien intelligence demands prostration from all devotees, demanding each devotee to divorce from their old morals, morale\ldots\ \ldots to be replaced with a new  cognizance, a new moral for judgment\ldots and all based on algorithmic decision making; essentially, the dawn of a new idol is upon us, and, to be sure, this new idol is as hollow as a clay figurine of old. 

The extreme, of course, is so\textendash called Big Data, which implies enough data or facts have been acquired to enable achieving Laplace's daemon. There is a worm of a lie at the heart of this argument: firstly, Laplace's argument never questioned the underlying principles upon which he would model the trajectories of some Avogadro number of particles, but he did most certainly question the human ability to ever track such a melee; secondly, reality is not the sum total of a set of facts, rather, reality is decided upon, but fools think of such narcissistic, solipsistic propositions as valid enough to be considered a theory. Reality does not become \emph{more real} with each additional ''fact,'' contrary, reality actually dissolves as knowledge is attained over a lifetime.

\centerline\noindent\begin{minipage}[tc]{0.75\textwidth}{
\begin{verse} 
\textbf{47}\\
Without stirring abroad\\
One can know the whole world;\\
Without looking out the window\\
One can see the way of heaven.\\
The further one goes\\
The less one knows.\\
Therefore the sage knows without having
to stir,\\
Identifies without having to see,\\
Accomplishes without having to act.\\
\vspace{7pt}
\hspace{2.5in}\textendash Tao de Ching \cite{tao}.
\end{verse}}
\end{minipage}

\vspace{12pt}

The principles of statistics provide an excellent lesson for those willing to listen to those well founded, solidly, mathematically defined precepts that speak of knowledge in terms of confidence, rather than certainties. But, all-to-often does a researcher perceive statistics a type of game, where the goal is to ''beat'' the test statistic, and, if the celebrated goal be achieved, the happy researcher makes a victory lap and claims to have discovered or proved something or another. Well, just because you ''beat'' the test statistic, doesn't mean you win the challenge. For example, if your calculated statistic be 2.0001 and compare this to a test statistic of 2.0\ldots should you conclude victory? Obviously not. One should be conservative in their approach to statistics, which would require a statistic at least an order higher than the test statistic\textemdash that would build confidence, eh. ''But, that would mean an obvious result,'' would be said in reply and I cannot deny this fact, for the application of statistics is not to discover the unknown, but rather to confirm what is already known.

If there be any bit of wisdom I can pass along to some youthful scientist, it would be thus: \emph{If you listen very closely, Sophia will tell you the truth, she will softly whisper in your ear, ''I don't know.''} Not the answer sought by a scientific mind, which fundamentally seeks \emph{to know} truth; but, truth is elusive, at best, and the world may be nothing more than a Rorschach inkblot, after-all.

\newpage

%Archangel Michael killing the dragon.

\centering
    \vfill
 \noindent
\hspace*{-\oddsidemargin}
    \vfill
   \centerline\noindent
   \makebox[0.75\textwidth]{\includegraphics[width=300pt]{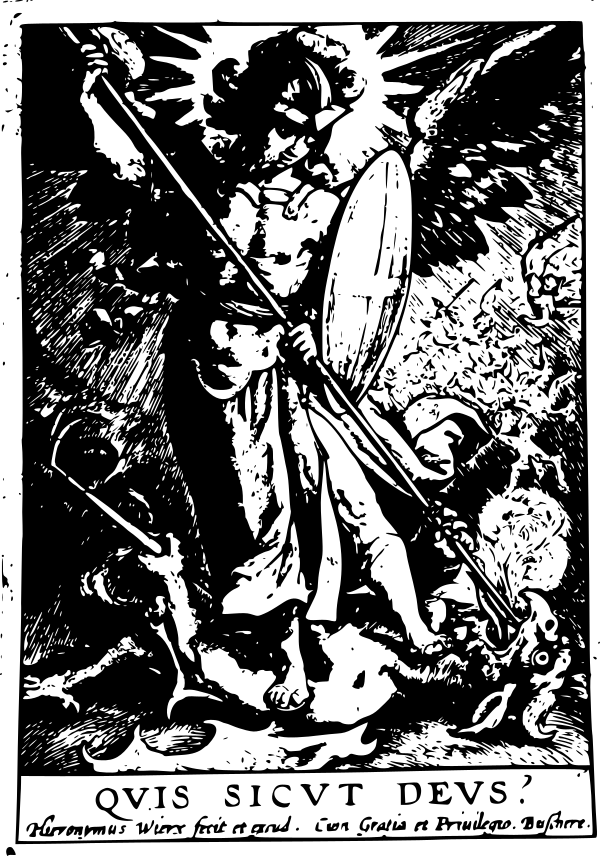}}
   
   \small Quis sicut Deus? "Who [is] like God?"
    \vfill    
    \vfill
    
\flushleft\footnotesize{
Title: ''The Archangel Michael slaying the dragon'', 

Artist: Hieronymus Wierix,

Date: 16th\textendash 17th century,

Department: Achenbach Foundation,

Accession Number: 1963.30.15032,

Credit Line: Achenbach Foundation for Graphic Arts,

\url{https://art.famsf.org/hieronymus-wierix/archangel-michael-slaying-dragon-19633015032}
}

\section{Appendix}\label{Appendix}

\begin{center}
 \begin{minipage}{1\textwidth}

\begin{example}[\textbf{Convolution property:} \emph{derivative, independent variable}]\label{TH1}
 
 If a derivative is applied over a convolution and the variable is independent of the convolution, then the derivative yields the summation of successive derivatives of each function involved in the convolution, \emph{viz.}:
 \begin{equation*}
  \frac{\partial}{\partial t}\Big(G\ast f\Big)=\Big(G'\ast f\Big) + \Big(f'\ast G\Big)
 \end{equation*}
\end{example}

\begin{example}[\textbf{Convolution property:} \emph{derivative, dependent variable}]\label{TH2}
 
 If the derivative happens to be with respect to the dependent variable, the derivative can be placed onto one singular function involved in the convolution, \emph{viz.}:
 \begin{equation*}
  \frac{\partial}{\partial x}\Big(G\ast f\Big)(x)=\Big(G'\ast f\Big)(x) = \Big(f'\ast G\Big)(x),
 \end{equation*}
 \noindent where symbol $(x)$ signifies what variable the convolution is over and prime ($'$) represents, in this case, a spatial derivative.
\end{example}

\begin{example}[\textbf{Convolution Theorem:}]\label{TH3}
 
 The Fourier transform ($\mathscr{F}$) of a convolution equals the multiplication of the transforms of each function, \emph{viz.}:
 \begin{equation*}
  \mathscr{F}\Big\{f_1\ast f_2\ast\cdots\ast f_n\Big\}(x)= \mathscr{F}\{f_1\}\mathscr{F}\{f_2\}\cdots\mathscr{F}\{f_n\}
 \end{equation*}
\end{example}

\centereqn{h}{\textwidth}{\vspace{7pt}\centering\noindent\rput[r](-3pt,3pt){\pgfornament[scale=.35]{72}}
\large{\; Formulae\;}%
\rput[l](3pt,3pt){\pgfornament[scale=.35]{73}}\\
\rput(0,0){\pgfornament[scale=.35]{85}}\\
\par\noindent\normalsize\flushleft

\vspace{7pt}
%\centering \rput[r](-3pt,3pt){\pgfornament[scale=.35]{72}}
%\large{\; Formulae\;}%
%\rput[l](3pt,3pt){\pgfornament[scale=.35]{73}}\\
%\rput(0,0){\pgfornament[scale=.35]{85}}\\
%\par\noindent\normalsize

\begin{equation}
 \frac{1}{b+D(2\pi s)^2}\supset \frac{e^{\sqrt{\frac{b}{D}}\, x} \theta (-x)}{2 \sqrt{Db} }+\frac{e^{-\sqrt{\frac{b}{D}}\, x} \theta (x)}{2 \sqrt{Db} }
\end{equation}  
   
\begin{equation}
\begin{split}
  & \frac{e^{-D(2\pi s)^2 t}}{b+D(2\pi s)^2}   \supset  \frac{e^{bt}e^{-\left(bx/\sqrt{Db}\right)}}{4\sqrt{Db}}\, \mathrm{\textit{erfc}}\left(\frac{2t\sqrt{Db}-x}{2\sqrt{Dt}}\right)+ \\
  &  \hspace{2in} \frac{
 e^{bt}e^{\left(bx/\sqrt{Db}\right)}}{4\sqrt{Db}}\, \mathrm{\textit{erfc}}\left(\frac{2t\sqrt{Db}+x}{2\sqrt{Dt}}\right)
 \end{split}
\end{equation}

Where \textit{erfc(x)} is the complementary error function, defined as $1-\textit{erf}(x)$.

\vspace{7pt}
}

\end{minipage}

\end{center}

\newpage 
%\medskip
\section{References}

\begingroup
\renewcommand{\section}[2]{}%

\end{document}